\documentclass[12pt,epsf]{article}
 \usepackage[compat=1.1.0]{tikz-feynman}
 \usepackage{verbatim,graphics,graphicx,color,slashed,amsmath}
 \usepackage{bm}
 \usepackage{cite}
 \usepackage{amssymb}
 \usepackage{braket}
 \usepackage{ascmac}
 \usepackage{multirow}
 \usepackage[compat=1.1.0]{tikz-feynman} 
\usepackage[]{hyperref}
 \hypersetup{colorlinks,bookmarks,unicode,linktocpage=true,linkcolor=blue, anchorcolor=blue, citecolor=blue}
 \usepackage{comment} 
\newcounter{num}

\begin{document}
\thispagestyle{empty}
\vspace*{-15mm}
\baselineskip 1pt
\begin{flushright}
\begin{tabular}{l}
\end{tabular}
\end{flushright}
\baselineskip 24pt
\vglue 10mm

\vspace{15mm}
\begin{center}
{\Large\bf Higgs-Portal Spin-1 Dark Matter\\ with Parity-Violating Interaction}
\vspace{7mm}

\baselineskip 18pt
{\bf Kimiko Yamashita}
\vspace{2mm}

{\it 
\, \, \, \, Department of Physics, Ibaraki University, Mito 310-8512, Japan
\newline \newline
kimiko.yamashita.nd93@vc.ibaraki.ac.jp}\\
\vspace{10mm}
\end{center}
\begin{center}
\begin{minipage}{14cm}
\baselineskip 16pt
\noindent
\begin{abstract}
We introduce the spin-1 $U(1)_X$ gauged field $X$ with $Z_2$ odd dark parity to evade the current strong constraints on kinetic mixing.
Then, $X$ becomes stable and a candidate for the dark matter.
The lowest mass dimension of interaction is six, and the type is the Higgs portal.
Two types of dim-$6$ operators are introduced. 
We consider the freeze-out dark matter scenario.
With the limit of null momentum transfer, a parity odd operator is free from the direct detection constraints.
Accordingly, the strong constraints on a parity even operator indicate turning on this parity odd operator to realize the dark matter relic density of the Universe.
With the $1$~TeV cut-off scale, our dark matter of around $400$~GeV mass can explain the dark matter relic density and is allowed from the LUX-ZEPLIN experiment of the direct detection.
\end{abstract}
\end{minipage}
\end{center}

\baselineskip 18pt
\def\thefootnote{\fnsymbol{footnote}}
\setcounter{footnote}{0}

\newpage

\section{Introduction}\label{sec:introduction}
Dark photons are known as extra $U(1)$ gauge bosons mirroring the abelian gauge bosons in the Standard Model (SM), 
i.e., photon, and well studied~\cite{Holdom:1985ag} (see \cite{Fabbrichesi:2020wbt} for a review). 
The dark photon interacts with the SM fermions through the kinetic mixing with the field strength of SM $U(1)_Y$ gauge boson.
For the low energy, the kinetic mixing between dark photon and $U(1)_Y$ gauge boson can be interpreted with 
the mixing between the dark photon and SM photon effectively after integrating out $Z$ boson~\cite{Fabbrichesi:2020wbt}.
The kinetic mixing parameter of dark photon and SM photon mixing is orders of $10^{-16}$--$10^{-5}$ depending on the dark photon mass
with the range of $10^{-17}$--$10^{5}$~eV well bellow the $Z$ boson mass~\cite{Caputo:2021eaa}.
For the TeV scale dark photon, the constraints on the mixing angle become weaker 
by strong suppression of the cross section by a small mixing angle to the fourth for the collider search, 
e.g., a cross section of process $pp \to X \to \ell \ell $ with dark photon $X$ and $\ell = e, \mu$.
Recalling an order nano barn cross section for process of $pp \to Z \to \ell \ell$ at the 13~TeV Large Hadron Collider (obtained with \verb|MadGraph5_aMC@NLO|~\cite{Alwall:2014hca}, 
applying typical experimental kinematic selections on the leptons of $p_T > 25$~GeV and $|\eta| < 2.4$),
the TeV scale dark photon production cross section should be enough smaller than the order of fb for the mixing angle with 0.01.

By reflecting the fact that the tiny/relatively small mixing angles are only allowed from the experimental/observational constraints,
it may be time to start considering another search strategy option for the dark photon by assuming some symmetry forbids the kinetic mixing of the dark photon to the SM sector.
If we assume that the dark photon has an odd parity under the $Z_2$ symmetry, whereas each SM particle has an even parity,
the kinetic mixing is forbidden, and leading interactions are from dimension-6 (dim-6) operators between dark photon and Higgs boson\footnote{If two different vector dark photons $X_1$ and $X_2$ are included, 
the dim-6 operators having three different field strengths, i.e., $B_\mu^{\ \nu} X_{1\nu}^{\ \alpha} X_{2\alpha}^{\ \ \mu}$ and $\tilde{B}_{\mu}^{\ \ \nu} X_{1\nu}^{\ \ \alpha} X_{2\alpha}^{\ \ \mu}$ exist. 
Here, $B^{\mu\nu}$ is the field strength of $U(1)_Y$ gauge boson, 
$X_1^{\mu\nu}$ and $X_2^{\mu\nu}$ are the field strengths of the two different dark photons $X_1$ and $X_2$, respectively. This case is studied in~\cite{Aebischer:2022wnl}.}.
As a result, this Higgs-portal dark photon is stable and can be a candidate for dark matter (DM).
We consider the freeze-out DM scenario.

The paper is organized as follows. In Section~\ref{sec:eft}, we list the relevant set of dim-$6$ effective operators in this study.
In Section~\ref{sec:dm_relic}, we show the model parameters to realize the current DM relic density in the Universe.
We study constraints from the direct detection in Sec.~\ref{sec:dm_direct}.
We summarize and conclude in Sec.~\ref{sec:summary}.

\section{Higgs-Portal Dark Photon Dark Matter in Dim-6}\label{sec:eft}
We consider the dark photon to be a gauge boson of the extra dark gauged $U(1)_X$ and dark parity odd, whereas we consider SM particles no dark charge and even dark parity.
In this case, the kinetic mixing is absent, and the dark photon can be a candidate for the DM in the Universe.
Interactions between dark photon and SM particles up to dim-6 are following the Higgs-portal dark photon operators:
\begin{align}
\mathcal{O} &= (H^{\dagger}H) X_{\mu\nu} X^{\mu\nu}, \label{eq:op1}\\
\mathcal{\tilde{O}} &= (H^{\dagger}H) \tilde{X}_{\mu\nu} X^{\mu\nu}, \label{eq:op2}\
\end{align}
where $H$ is the SM Higgs doublet, the field strength of vector DM $X^\mu$ is $X^{\mu\nu} = \partial^\mu X^\nu - \partial^\nu X^\mu$, 
and the dual field strength is $\tilde{X}_{\mu\nu} = \frac{1}{2} \epsilon_{\mu\nu\rho\sigma} X^{\rho\sigma} $. 
Here, $\epsilon_{\mu\nu\rho\sigma}$ is the totally antisymmetric Levi-Civita tensor with $\epsilon_{0123} = 1$.
The  Lagrangian for the Higgs-portal vector DM is:
\begin{align}
\mathcal{L_\mathrm{VDM}} = \frac{m^2_{X}}{2} X^\mu X_{\mu} + \frac{C}{\Lambda^2} \mathcal{O} +  \frac{\tilde{C}}{\Lambda^2} \mathcal{\tilde{O}},
\label{eq:lag}
\end{align}
where $C$ and $\tilde{C}$ are the Wilson coefficients for the corresponding operator $O$ and $\tilde{O}$ in Eqs.~\eqref{eq:op1} and  \eqref{eq:op2} respectively and $\Lambda$ is the characteristic scale of new physics,
i.e., around the mass scale of the mediator between the vector DM and Higgs boson.
Note that $\tilde{O}$ gives odd under the parity and Time reversal transformations respectively due to $\epsilon_{\mu\nu\rho\sigma}$.

We include vector DM mass in the Lagrangian~\eqref{eq:lag}. For the UV physics, it can be considered that the vector DM mass 
appears by the spontaneous symmetry breaking of dark gauge $U(1)_X$ in the dark sector with a dark charged scalar boson taking a vacuum expectation value.
Introducing the odd dark parity for the vector DM may require other mechanisms, but in this work, 
we do not consider such microscopic dynamics in detail as we consider the effective field theory or low energy theory for the vector DM.

\section{Dark Matter Relic Density}\label{sec:dm_relic}
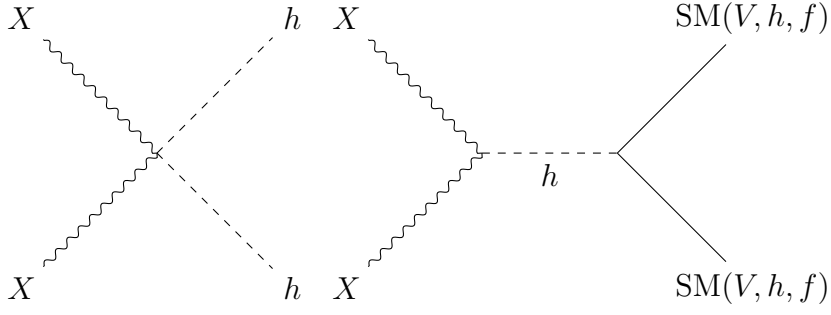
\begin{figure}[!t]
\begin{center}
	\begin{tikzpicture}[baseline=($(int)$)]
		\begin{feynman}[inline=($(int)$),medium]
			\vertex (in1) at (-1.8, 1.8) {\( X \)};
			\vertex (in2) at (-1.8, -1.8) {\( X \)};
			\vertex (int) at (0, 0);
			\vertex (out1) at (1.8, 1.8) {\( h \)};
			\vertex (out2) at (1.8, -1.8) {\( h \)};
		
			\diagram* {
				(in1) -- [photon] (int) -- [scalar] (out2), 
				(in2) -- [photon] (int) -- [scalar] (out1), 
			};
		\end{feynman}
	\end{tikzpicture}		
		\begin{tikzpicture}[baseline=($(int)$)]
		\begin{feynman}[inline=($(int)$),medium]
			\vertex (in1) at (-1.8, 1.8) {\( X \)};
			\vertex (in2) at (-1.8, -1.8) {\( X \)};
			\vertex (int) at (0, 0);  
			\vertex (out) at (1.8, 0);
			\vertex (out1) at (3.6, 1.8) {\( \mathrm{SM} (V, h, f) \)};
			\vertex (out2) at (3.6, -1.8) {\( \mathrm{SM} (V, h, f) \)};
		
			\diagram* {
				(in1) -- [photon] (int) -- [scalar] (int), 
				(in2) -- [photon] (int) -- [scalar] (int), 
				(int) -- [scalar, edge label'=\(h\)] (out),
				(out) --[scalar] (out) -- (out1),
				(out) --[scalar] (out) -- (out2)
			};
		\end{feynman}
	\end{tikzpicture}	
	\end{center}
\caption{Feynman diagrams for dark matter annihilation}
\label{fig:diagram_RelicD}
\vspace{5mm}
\end{figure}

As a first step, we consider DM the Weekly Interacting Massive Particle (WIMP), produced via the freeze-out mechanism in this paper.
Studying the Feebly interacting DM with non-thermal production, i.e., the freeze-in mechanism is future work.
For deriving the vector DM relic density from the Higgs-portal dim-6 interactions, annihilation channels are
$X X \to hh, VV$, and $f{\bar f}$ with $h$ being the SM Higgs boson,  $V=W, Z$, and $f$ being the SM fermions as shown in Fig.~\ref{fig:diagram_RelicD}.  
We have omitted $t$-channel contribution on $X X \to hh$ as it is suppressed by $1/\Lambda^4$ for the amplitude level.
The number density for the DM $n_X$ follows the Boltzmann equation:
\begin{align}
\dot{n}_{X} + 3Hn_{X} &= - \langle\sigma v_\mathrm{rel}\rangle_\mathrm{eff}
\left(n^2_X - (n^\mathrm{eq}_X)^2\right),
\label{eq:boltzmann}
\end{align}
where
\begin{align}
\langle\sigma v_\mathrm{rel}\rangle_\mathrm{eff} &= 2\langle\sigma v_\mathrm{rel}\rangle_{X X \to h h}
+ 2\langle\sigma v_\mathrm{rel}\rangle_{X X \to W^+ W^-}
+ 2\langle\sigma v_\mathrm{rel}\rangle_{X X \to ZZ } +2\langle\sigma v_\mathrm{rel}\rangle_{X X \to f{\bar f}}.
\end{align}
$n^\mathrm{eq}_X$ is the DM number density in thermal equilibrium. 
DM decoupling is done when it is non-relativistic, so the DM relative velocity squared $v^2_\mathrm{rel}$ at that time, i.e., freezing out, is enough smaller 1.
In this period, the thermal averaged annihilation cross sections $\langle\sigma v_\mathrm{rel}\rangle$ can be written as~\cite{Mambrini2021}
\begin{align}
\langle\sigma v_\mathrm{rel}\rangle_{X X \to i j} &= a_{ij} + \frac{6b_{ij}}{x}, 
\end{align}
where
\begin{align}
\sigma v_\mathrm{rel} (X X \to i j) &= a_{ij} + b_{ij} v^2_\mathrm{rel},\label{eq:xsec_v2}\\
x&=m_X/T.
\end{align}
Here, $(i,j)$ denotes $(h,h)$, $(W^+, W^-)$, $(Z, Z)$, and $(f,{\bar f})$, respectively. 
$T$ is the temperature of the thermal bath.
The $s$-wave contribution is dominant for the annihilation cross sections in the limit $v_\mathrm{rel} \to 0$.
The $p$-wave contribution $(\propto v^2_\mathrm{rel}$ in Eq.~\eqref{eq:xsec_v2}) is considered when the leading term of $s$-wave vanishes.
For obtaining the cross section up to $v^2_\mathrm{rel}$ (Eq.~\eqref{eq:xsec_v2}), we expand the cross section with $v^2_\mathrm{rel}$ and obtain the following form as
\begin{align}
\sigma v_\mathrm{rel} (X X \to i j) = \frac{|\mathcal{M}_{X X \to ij}|^2}{32\pi m^2_X}\sqrt{1-\frac{m^2_i}{m^2_X}}\left(1-\frac{v^2_\mathrm{rel}}{4} \left(1+\frac{1}{2(1-m^2_i/m^2_X)}\right)\right),
\end{align}
where $m_i$ is $i$ field mass, the squared matrix element (up to $v^2_\mathrm{rel}$) $|\mathcal{M}_{X X \to ij}|^2$ includes the symmetric factor 
for identical particles in initial states (i.e., a DM particle pair $X X$) and final states (i.e., $h h$ and $Z Z$). 
The spin states are averaged/summed up for the initial/final state of the spinning particle, i.e., the vector boson and fermion.
When 
\begin{align}
|\mathcal{M}_{X X \to ij}|^2 = a'_{ij} + b'_{ij} v^2_\mathrm{rel},
\end{align}
\begin{align}
\sigma v_\mathrm{rel} (X X \to i j) &=  \frac{1}{32\pi m^2_X}\sqrt{1-\frac{m^2_i}{m^2_X}}\left( a'_{ij}+v^2_\mathrm{rel}\left( b'_{ij} - \frac{a'_{ij}}{4} \left(1+\frac{1}{2(1-m^2_i/m^2_X)}\right) \right)\right).
\end{align}
We obtain the squared scattering matrix elements for final states by implementing the model into FeynRules~\cite{Christensen:2008py,Alloul:2013bka,Christensen:2009jx}
and using CalcHEP~\cite{Belyaev:2012qa}.
The squared scattering matrix elements up to $v^2_\mathrm{rel}$ are
\begin{align}
|\mathcal{M}_{X X \to h h}|^2 &= \frac{16 C^2 m_X^4 (2 m_X^2 + m_h^2)^2}{3 \Lambda^4 (4 m_X^2 - m_h^2)^2}\nonumber \\
&\, \, \, \,  + \frac{16 m_X^8 (32 m_X^2 (C^2+\tilde{C}^2) + 6 m_h^2  (C^2+4 \tilde{C}^2))}{9 \Lambda^4 (4 m_X^2 - m_h^2)^3}v^2_\mathrm{rel}\nonumber \\
&\, \, \, \,  - \frac{16 m_X^4 m_h^4(9 C^2 m_X^2   + 2 m_h^2 (C^2+\tilde{C}^2))}{9 \Lambda^4 (4 m_X^2 - m_h^2)^3}v^2_\mathrm{rel}, \label{eq:XX_hh}\\ 
|\mathcal{M}_{X X \to W^+ W^-}|^2 &= \frac{32 \pi^2  \alpha^2 m^4_X v^4 C^2 \left(4m^4_X - 4 m^2_X m^2_W + 3m^4_W \right)}{3 \Lambda^4 m^4_W s^4_W(4 m^2_X- m^2_h)^2}\nonumber\\
&\, \, \, \, + \frac{8 \pi ^2 \alpha^2 m^2_X v^4 (C^2+\tilde{C}^2) (2 m^2_X - 2 m^2_W + m^2_h)}{9 \Lambda^4 m^4_W s^4_W} v^2_\mathrm{rel}, \label{eq:XX_ww}\\
|\mathcal{M}_{X X \to Z Z}|^2 &=\frac{1}{2} |\mathcal{M}_{X X \to W^+W^-}|^2 (m_W\rightarrow m_Z, s_W\rightarrow s_W c_W), \label{eq:XX_zz}\\
|\mathcal{M}_{X X \to f {\bar f}}|^2 &= \frac{64 C^2 m_X^4 m^2_f (m_X^2 - m^2_f)}{\Lambda^4 (4m_X^2 - m^2_h)^2} + \frac{8 m^2_X m^2_f (C^2+\tilde{C}^2)}{3 \Lambda^4} v^2_\mathrm{rel}, \label{eq:XX_ff}
\end{align}
where $c_W = \cos{\theta_W}$, $s_W = \sin{\theta_W}$, $\theta_W$ is the Weinberg angle, 
$\alpha$ is the fine structure constant, and $v$ is the vacuum expectation value of the SM Higgs field.
Here, $f{\bar f} = t{\bar t}, b{\bar b}$ and $m_f = m_t, m_b$  are top and bottom quark masses. 
$s$-wave dominates all the annihilation channels for the $\mathcal{O}$ operator, whereas $p$-wave suppressed for the $\mathcal{\tilde{O}}$ operator.

By solving the Boltzmann equation Eq.~\eqref{eq:boltzmann}, we can obtain the abundance at present $Y_\mathrm{DM} = n_X/s$ that related to the relic density 
for the vector DM as follows:
\begin{align}
\Omega_\mathrm{DM} h^2 = 0.2744 \left(\frac{Y_\mathrm{DM}}{10^{-11}}\right)\left(\frac{m_X}{100~\mathrm{GeV}}\right).
\end{align}
\begin{figure}[t!]\center
\includegraphics[width=0.495\textwidth,clip]{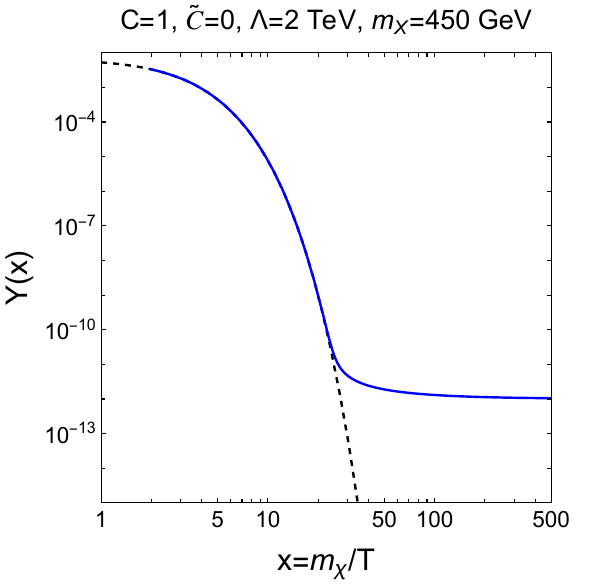}
\includegraphics[width=0.495\textwidth,clip]{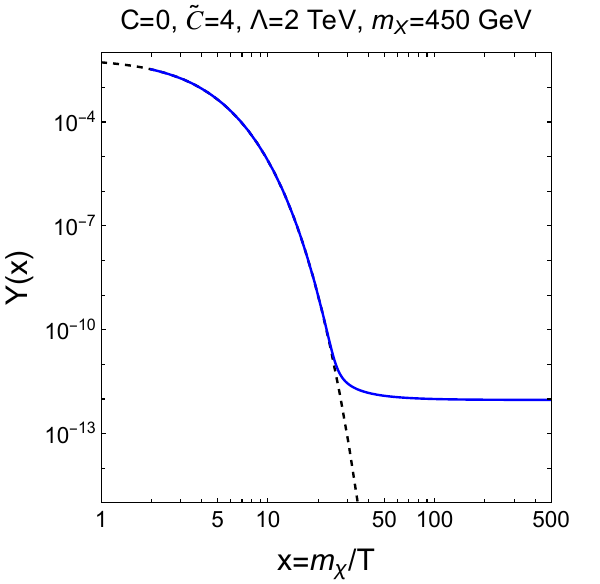}\\
\includegraphics[width=0.495\textwidth,clip]{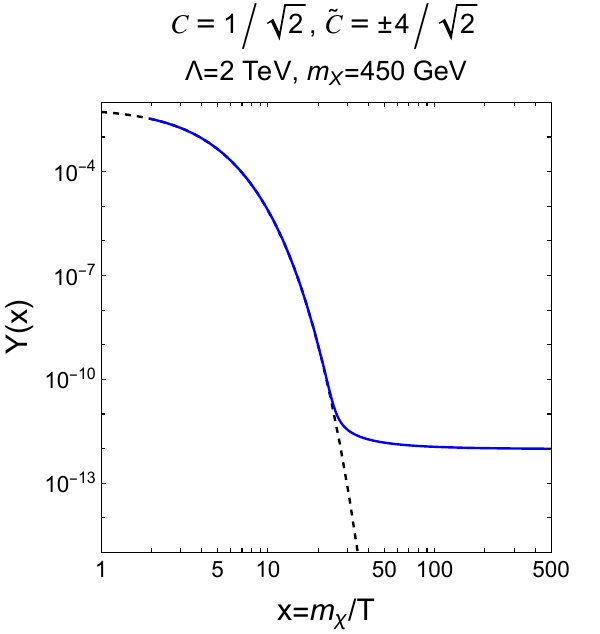}
\caption{The relic abundance for dark matter as a function of $x=m_X/T$ as the solution of Boltzmann equation (blue solid line),
compared to the thermal equilibrium abundance (black dashed line). We take dark matter mass $m_X = 450$~GeV. 
For the relic abundance shown as a blue solid line,
we take $(C, \tilde{C}) = (1, 0)$ (upper left), $(0, 4)$ (upper right),  $(1/\sqrt{2}, \pm 4/\sqrt{2})$ (bottom) with $\Lambda = 2~$TeV.}
\label{fig:evol_Y}
\vspace{5mm}
\end{figure}
Figure~\ref{fig:evol_Y} shows the relic abundance for DM as a function of $x=m_X/T$ in a blue solid line.
We take DM mass $m_X = 450$~GeV. 
For the relic abundance shown as a blue solid line,
we take $(C, \tilde{C}) = (1, 0), (0, 4), (1/\sqrt{2}, \pm 4/\sqrt{2})$ with $\Lambda = 2~$TeV respectively.
For reference, the black dashed line is depicted as the thermal equilibrium abundance.
We obtain the relic density of the current Universe $\Omega_\mathrm{DM} h^2 \sim 0.1$ with these parameter sets.

An approximate analytic solution to the Boltzmann equation exists~\cite{Scherrer:1985zt,Cline:2013gha,Mambrini2021}.
The present abundance is estimated as follows:
\begin{align}
Y_\mathrm{DM} \simeq \sqrt{\frac{45g_*}{\pi g^2_{*s}}}\frac{x_f}{m_X\sqrt{8\pi}\overline{M}_\mathrm{pl}\sum_{ij}2(a_{ij}+3b_{ij}/x_f)},\label{eq:YDM_analytic}
\end{align}
where 
\begin{align}
x_f &\sim \ln\left(c(c+2)\sqrt{\frac{45}{8}}\frac{1}{2\pi^3}\frac{g_X\overline{M}_{pl}m_X\sum_{ij}2(a_{ij}+6b_{ij}/x_f)}{\sqrt{x_f}\sqrt{g_{*s}}}\right)
\end{align}
$\overline{M}_\mathrm{pl} = M_\mathrm{pl}/\sqrt{8\pi} =  2.43\times10^{18}~$GeV is the Reduced Planck mass with the Planck mass
$M_\mathrm{pl} = 1.22\times10^{19}~$GeV, and $g_{*s}$ and $g_{*} $ are the effective numbers of the relativistic degree of freedom in entropy and radiation, respectively.
$g_X = 3$ is the degree of freedom for the vector DM. In $\sum_{ij}$, each final state of the annihilation processes is summed up.
$c$ is an order unity parameter determined numerically by solving the Boltzmann equation and is set equal to $0.5$ for this analytical approximation~\cite{Mambrini2021}.

\begin{figure}[t!]\center
\includegraphics[width=0.495\textwidth,clip]{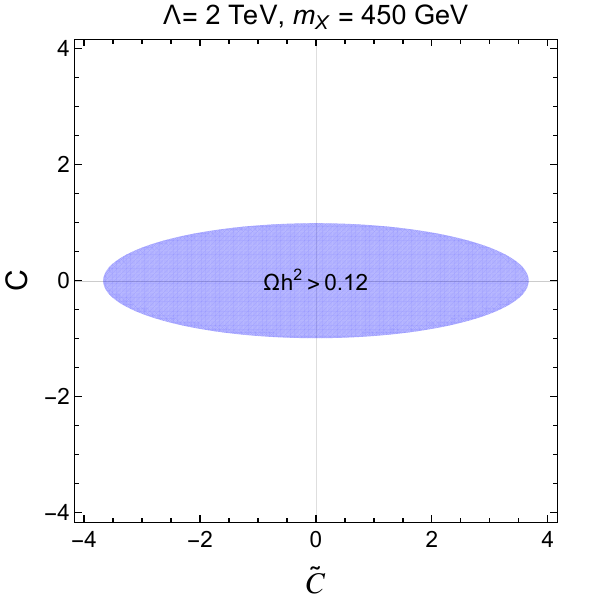}\\
\includegraphics[width=0.495\textwidth,clip]{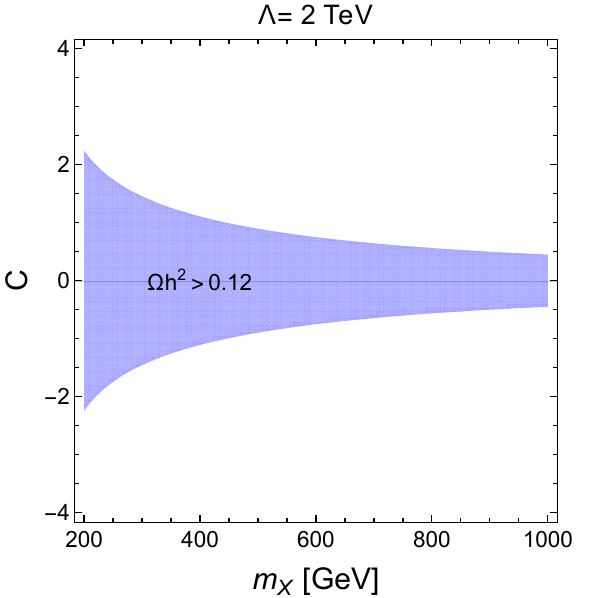}
\includegraphics[width=0.495\textwidth,clip]{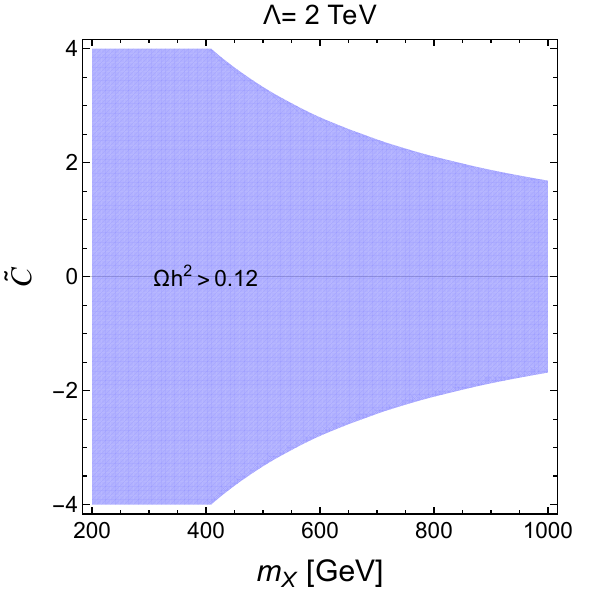} 
\caption{
Upper: Parameter space for $C$ vs $\tilde{C}$, satisfying relic density. We take  $m_X = 450\,{\rm GeV}$.
Lower: The relic abundance in the parameter spaces of $(m_X, C)$ with $\tilde{C} = 0$ (left) and  $(m_X, \tilde{C})$ with $C = 0$ (right).
The relic abundance for dark matter is overproduced in blue regions, namely, $\Omega h^2>0.12$, and it saturates the observed value along the boundary of the blue region.
We take $\Lambda = 2$~ TeV.}
\label{fig:RelicD}
\vspace{5mm}
\end{figure}
By utilizing the approximate analytic formula Eq.~\eqref{eq:YDM_analytic} with $x_f = 23$, we depict a contour plot for the relic abundance in the parameter space
of $(C, \tilde{C})$ with $m_X = 450\, {\rm GeV}$ (upper), $(m_X, C)$ with $\tilde{C} = 0$ (lower left) and  $(m_X, \tilde{C})$ with $C = 0$ (lower right) respectively in Fig.~\ref{fig:RelicD}. 
We take $\Lambda = 2~$TeV.
The relic abundance for dark matter is overproduced in blue regions, namely, $\Omega h^2>0.12$, and it saturates the observed value along the boundary of the blue region.

\section{Direct Detection Constraints}\label{sec:dm_direct}
\begin{figure}[t!]
\begin{center}
	\begin{tikzpicture}[baseline=($0.5*(int1)+0.5*(int3)$)]
		\begin{feynman}[inline=($0.5*(int1)+0.5*(int3)$),medium]

			\vertex (in1) at (-1.95, 2.1) {\( X \)};
			\vertex (in2) at (-1.95, -2.1) {\( q \)};
			\vertex (int1) at (0, 0.9);
			\vertex (int3) at (0, -0.9);
			\vertex (out1) at (1.95, 2.1) {\( X \)};
			\vertex (out2) at (1.95, -2.1) {\( q \)};
		
			\diagram* {
				(in1) -- [photon] (int1) -- [photon] (out1),
				(in2) -- [fermion] (int3) -- [fermion] (out2),
				(int1) -- [scalar, edge label = \( h \)] (int3)
			};
		\end{feynman}
	\end{tikzpicture}
	
	\end{center}
\caption{Feynman diagram for dark matter direct detection}
\label{fig:diagram_DD}
\vspace{5mm}
\end{figure}
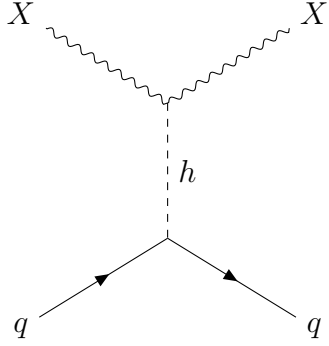

For the spin-independent elastic scattering between DM and nucleons, 
we first obtain the following effective Lagrangian for DM and quarks due to the Higgs exchange as shown in Fig.~\ref{fig:diagram_DD},
\begin{align}
\mathcal{L}_{X, q} = - 2\frac{m_q}{m^2_h\Lambda^2} \bar{q}{q}(C X_{\mu\nu} X^{\mu\nu} + \tilde{C} \tilde{X}_{\mu\nu} X^{\mu\nu})
\end{align}
The resulting effective Lagrangian between the DM $X$ and nucleon $N (=p, n)$ is as follows
\begin{align}
\mathcal{L}_{X, N} &=  - 2\frac{m_N}{m^2_h\Lambda^2}  \sum_{q=\mathrm{all}} f^N_q \bar{N}N (C X_{\mu\nu} X^{\mu\nu} + \tilde{C} \tilde{X}_{\mu\nu} X^{\mu\nu}) \label{eq:Leff} \\
	&=  - 2\frac{m_N}{m^2_h\Lambda^2}  \left(\frac{7}{9}\sum_{q=u,d,s}  f^N_q +\frac{2}{9} \right) \bar{N}N (C X_{\mu\nu} X^{\mu\nu} + \tilde{C} \tilde{X}_{\mu\nu} X^{\mu\nu})\label{eq:Leff2},
\end{align}
where
\begin{align}
f^N_q = \frac{m_q}{m_N} \Braket{N| \bar{q}q|N}.\label{eq:fraction}
\end{align}
Here, $m_N$ is the mass of proton or neutron, $m_N \equiv m_p \sim m_n$ and is approximately the same in our analyses for the direct detection constraints,
$m_q (q= u, d, c, s, t, b)$ is the mass of the quark.
$f^N_q$ is the mass fraction of quark $q$ inside the nucleon and its numerical values are $f^p_{u}=0.0208\pm 0.0015$ and $f^p_{d}=0.0411\pm 0.0028$ for a proton, 
$f^n_{u}=0.0189\pm 0.0014$ and $f^n_{d}=0.0451\pm 0.0027$ for a neutron~\cite{Hoferichter:2015dsa},  and 
$f^{p,n}_{s}=0.043\pm 0.011$ for both proton and neutron~\cite{Junnarkar:2013ac}.
Heavy quarks contributions to Eq.~\eqref{eq:fraction}  are obtained by computing the trace of the energy-momentum tensor in the nucleon state (from Eq.~\eqref{eq:Leff} to Eq.~\eqref{eq:Leff2}) ~\cite{Shifman:1978zn,Mambrini2021}.

From the effective Lagrangian~Eq.\eqref{eq:Leff2}, we compute the amplitude square $|\mathcal{M}_N|^2$ and cross section as
\begin{align}
 |\mathcal{M}_N|^2 &= \frac{64 C^2 m^4_N m^4_X}{\Lambda^4 m^4_h}\left(\frac{7}{9}\sum_{q=u,d,s}  f^N_q +\frac{2}{9} \right)^2, \\
 \sigma^{\mathrm{SI}}_{X,N}  &= \frac{|\mathcal{M}_N|^2}{16\pi(m_N+m_X)^2}\\
 	&= \frac{\mu^2_N}{\pi}\frac{4 C^2 m^2_N m^2_X}{\Lambda^4 m^4_h} \left(\frac{7}{9}\sum_{q=u,d,s}  f^N_q +\frac{2}{9}\right)^2,
\end{align}
where $\mu_N$ is the reduced mass of the DM and nucleon system, $\mu_N = m_X m_N/(m_X + m_N)$.
Note that there is no contribution from the parity odd operator in the limit of $t=0$ (null momentum transfer).
Finally, the corresponding spin-independent cross section between the DM $X$ and nucleus $N'$ is
\begin{align}
 \sigma^{\mathrm{SI}}_{X,N'} &= \frac{\mu^2_{N'}}{\pi}\frac{4 C^2 m^2_N m^2_X}{\Lambda^4 m^4_h A^2} 
\left(Z \left(\frac{7}{9}\sum_{q=u,d,s}  f^p_q +\frac{2}{9}\right) + (A-Z) \left(\frac{7}{9}\sum_{q=u,d,s}  f^n_q +\frac{2}{9}\right)\right)^2,
\end{align}
where $Z$ and $A$ are the proton and atomic numbers of the nucleus and $\mu_{N'}$ is the reduced mass of the DM and nucleus system, $\mu_{N'} = m_X m_{N'}/(m_X + m_{N'})$.

For comparing different experiments, recasting to the interacting cross section $\sigma^N$ of the DM and nucleon $N$ is used as
\begin{align}
\sigma^N = \frac{\mu^2_N}{\mu^2_{N'} A^2} \sigma^{\mathrm{SI}}_{X, N'}.
\end{align}

\begin{figure}[t!]\center
\includegraphics[width=0.7\textwidth,clip]{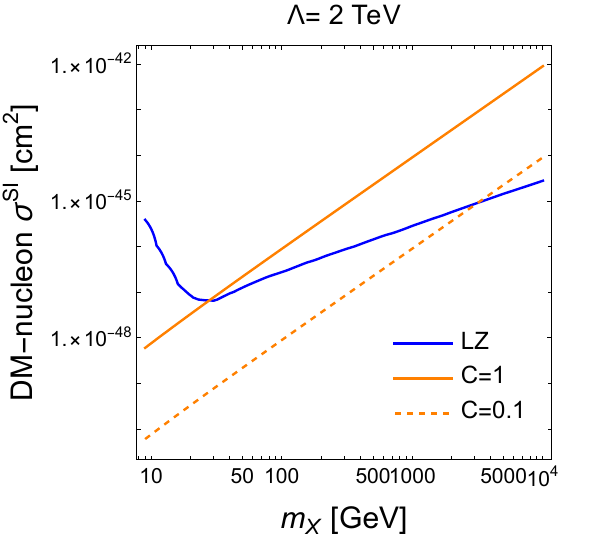}
\caption{The spin-independent cross section between the DM and nucleon as a function of the DM mass $m_X$,
compared to the constraints (blue solid, upper bounds) from the LZ experiment~\cite{LZ:2022lsv}. We take $\Lambda = 2~$TeV with $C = 1$ (orange solid) and $C= 0.1$ (orange dashed).}
\label{fig:DD_DM}
\end{figure}
The direct detection bound from the LUX-ZEPLIN (LZ) experiment~\cite{LZ:2022lsv} is tight as shown in Fig.~\ref{fig:DD_DM}.
We show the spin-independent cross section between the DM and nucleon as a function of the DM mass $m_X$, compared to the constraints (blue solid, upper bounds) 
from the LZ experiment~\cite{LZ:2022lsv}.
Here, we take $\Lambda = 2~$TeV with $C = 1$ (orange solid) and $C = 0.1$ (orange dashed).
\begin{figure}[t!]\center
\includegraphics[width=0.495\textwidth,clip]{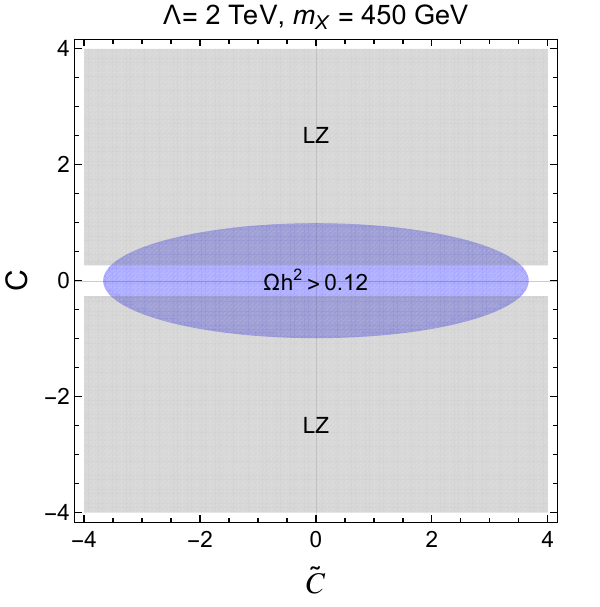}\\
\includegraphics[width=0.495\textwidth,clip]{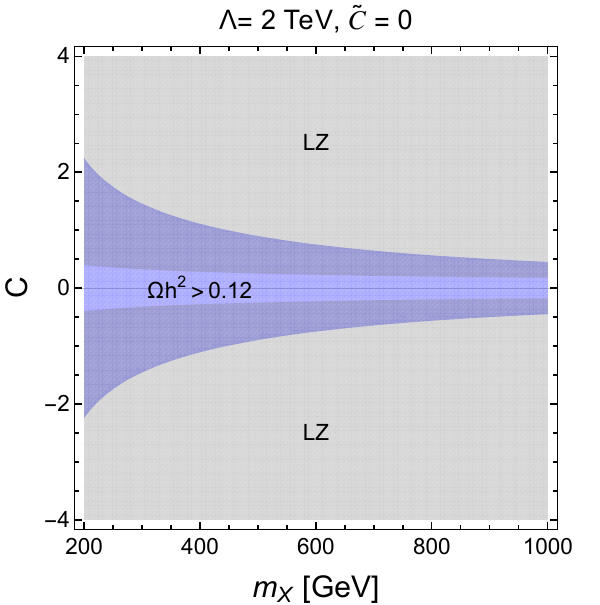}
\includegraphics[width=0.495\textwidth,clip]{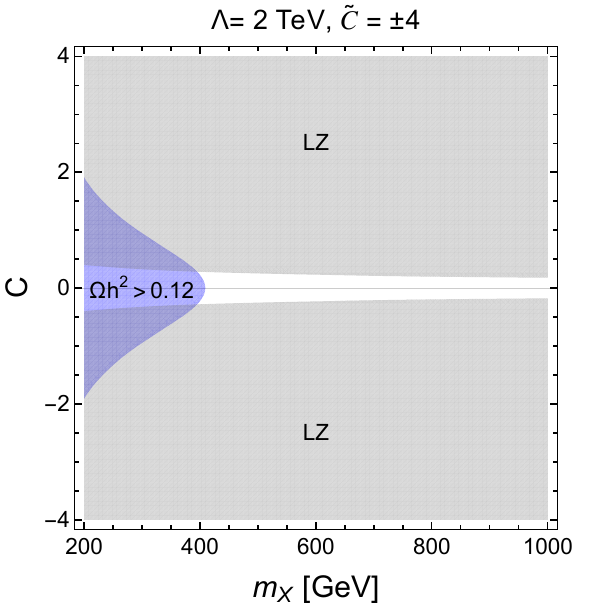}
\caption{
Upper: Parameter space for $C$ vs $\tilde{C}$. We take  $m_X = 450\,{\rm GeV}$.
Lower: Parameter spaces of $(m_X, C)$ with $\tilde{C} = 0$ (left) and $\tilde{C} = \pm4$ (right).
The relic abundance for DM is overproduced in blue regions, namely, $\Omega h^2>0.12$, and it saturates the observed value along the boundary of the blue region.
The gray regions are excluded by the LZ experiment~\cite{LZ:2022lsv}. We take $\Lambda = 2$~ TeV.}
\label{fig:DD_DM2}
\vspace{5mm}
\end{figure}
In Fig.~\ref{fig:DD_DM2}, we show the parameter spaces of $C$ vs $\tilde{C}$ with $m_X = 450\,{\rm GeV}$ (upper)
and  $(m_X, C)$ with $\tilde{C} = 0$ (lower left) and $\tilde{C} = \pm4$ (lower right). We take $\Lambda = 2$ TeV.
The present relic density for dark matter is beyond the observed value in the blue region, i.e., $\Omega h^2 > 0.12$.
The gray regions are excluded by the LZ experiment~\cite{LZ:2022lsv}.
The strong constraints from the direct detection measurement on the parity even operator 
indicate turning on the parity odd operator to realize the DM relic density of the Universe with $m_X < 1$~TeV.
As shown lower right of Fig.~\ref{fig:DD_DM2}, with $\tilde{C} = \pm4$ and $\Lambda = 2$~TeV (or $\Lambda/\sqrt{|\tilde{C}|} = 1$~TeV), 
our DM of $m_X \sim 400$~GeV can explain the DM relic density and allowed from the LZ experiment.

We can translate the mass of a new heavy resonance of $M \geq 1$~TeV to the Wilson coefficients of the dim-6 operators~\cite{Boughezal:2022nof} by
\begin{align}
\frac{\Lambda}{(\sqrt{|C_6|})} \geq \frac{1~\mathrm{TeV}}{g},
\end{align}
where $C_6$ is the Wilson coefficient of a dim-6 operator, and $g$ is the coupling of the heavy resonance.
Then, if we take $g = \sqrt{4\pi}$ at maximum and $\Lambda = 2$~TeV,
we have $|C_6| \leq 16\pi$.
Thus, it should be an acceptable DM effective field theory scenario having $\tilde{|C|} = 4$ with $\Lambda = 2$~TeV.
%

\section{Summary}\label{sec:summary}
We introduce the spin-1 $U(1)_X$ gauged field $X$ with $Z_2$ odd dark parity
for evading the current strong constraints on the kinetic mixing.
Then, $X$ becomes stable and a candidate for the dark matter.
The lowest mass dimension of interaction is six, and the type is the Higgs portal.
Two types of operators exist with the forms of $X^{\mu\nu}X_{\mu\nu}H^{\dagger}H$ (parity even) and $X^{\mu\nu}\tilde{X}_{\mu\nu}H^{\dagger}H$ (parity odd).

We investigate this DM model by checking the DM relic density and direct detection constraint.
For the relic density, the parity even operator contributes $s$-wave, whereas the parity odd operator contributes $p$-wave (suppressed).
With the limit of null momentum transfer, the non-zero Wilson coefficients from the parity odd operator are free from the direct detection constraints.
Interestingly, the strong constraints from the direct detection measurement on the parity even operator 
indicate turning on the parity odd operator to realize the DM relic density of the Universe with $m_X < 1$~TeV.
With $\Lambda/\sqrt{|\tilde{C}|} = 2~\mathrm{TeV}/\sqrt{4} = 1$~TeV, our DM of $m_X \sim 400$~GeV can explain the DM relic density and is allowed from the LZ experiment.

As DM can annihilate into $f\bar{f}$, $WW$, $ZZ$, and $hh$, there are cosmic ray signals that can be targeted from the indirect detection experiments.

For UV physics, the vector DM mass appears by the spontaneous symmetry breaking of dark gauge $U(1)_X$ in the dark sector with a dark charged scalar boson taking a vacuum expectation value. Introducing the odd dark parity for the vector DM may require other mechanisms, and we leave this for future work.

Investigating the non-thermal production of feebly interacting DM is also an interesting avenue for future research.
In this case, the DM production in the Higgs potential during the reheating epoch can be considered the initial condition of the feebly interacting DM density.
Our Higgs-portal vector DM model can be embedded in the Higgs inflation framework.
\paragraph*{Acknowledgements}
We have used the package TikZ-Feynman~\cite{Ellis:2016jkw} to draw the Feynman diagrams.
This work is supported by JSPS KAKENHI Grant Number JP24K17040.


\begin{thebibliography}{99}
\setlength{\itemsep}{0pt}
\bibitem{Holdom:1985ag}
B.~Holdom,
Phys. Lett. B \textbf{166}, 196-198 (1986)
doi:10.1016/0370-2693(86)91377-8

\bibitem{Fabbrichesi:2020wbt}
M.~Fabbrichesi, E.~Gabrielli and G.~Lanfranchi,
doi:10.1007/978-3-030-62519-1
[arXiv:2005.01515 [hep-ph]].

\bibitem{Caputo:2021eaa}
A.~Caputo, A.~J.~Millar, C.~A.~J.~O'Hare and E.~Vitagliano,
Phys. Rev. D \textbf{104}, no.9, 095029 (2021)
doi:10.1103/PhysRevD.104.095029
[arXiv:2105.04565 [hep-ph]].

\bibitem{Alwall:2014hca}
J.~Alwall, R.~Frederix, S.~Frixione, V.~Hirschi, F.~Maltoni, O.~Mattelaer, H.~S.~Shao, T.~Stelzer, P.~Torrielli and M.~Zaro,
JHEP \textbf{07}, 079 (2014)
doi:10.1007/JHEP07(2014)079
[arXiv:1405.0301 [hep-ph]].

\bibitem{Aebischer:2022wnl}
J.~Aebischer, W.~Altmannshofer, E.~E.~Jenkins and A.~V.~Manohar,
JHEP \textbf{06}, 086 (2022)
doi:10.1007/JHEP06(2022)086
[arXiv:2202.06968 [hep-ph]].

\bibitem{Mambrini2021}
Y. Mambrini, Particles in the dark Universe, Springer Cham (2021),
https://doi.org/10.1007/978-3-030-78139-2.

\bibitem{Christensen:2008py}
N.~D.~Christensen and C.~Duhr,
Comput. Phys. Commun. \textbf{180}, 1614-1641 (2009)
doi:10.1016/j.cpc.2009.02.018
[arXiv:0806.4194 [hep-ph]].

\bibitem{Alloul:2013bka}
A.~Alloul, N.~D.~Christensen, C.~Degrande, C.~Duhr and B.~Fuks,
Comput. Phys. Commun. \textbf{185}, 2250-2300 (2014)
doi:10.1016/j.cpc.2014.04.012
[arXiv:1310.1921 [hep-ph]].

\bibitem{Christensen:2009jx}
N.~D.~Christensen, P.~de Aquino, C.~Degrande, C.~Duhr, B.~Fuks, M.~Herquet, F.~Maltoni and S.~Schumann,
Eur. Phys. J. C \textbf{71}, 1541 (2011)
doi:10.1140/epjc/s10052-011-1541-5
[arXiv:0906.2474 [hep-ph]].

\bibitem{Belyaev:2012qa}
A.~Belyaev, N.~D.~Christensen and A.~Pukhov,
Comput. Phys. Commun. \textbf{184}, 1729-1769 (2013)
doi:10.1016/j.cpc.2013.01.014
[arXiv:1207.6082 [hep-ph]].

\bibitem{Scherrer:1985zt}
R.~J.~Scherrer and M.~S.~Turner,
Phys. Rev. D \textbf{33}, 1585 (1986)
[erratum: Phys. Rev. D \textbf{34}, 3263 (1986)]
doi:10.1103/PhysRevD.33.1585

\bibitem{Cline:2013gha}
J.~M.~Cline, K.~Kainulainen, P.~Scott and C.~Weniger,
Phys. Rev. D \textbf{88}, 055025 (2013)
[erratum: Phys. Rev. D \textbf{92}, no.3, 039906 (2015)]
doi:10.1103/PhysRevD.88.055025
[arXiv:1306.4710 [hep-ph]].

\bibitem{Hoferichter:2015dsa}
M.~Hoferichter, J.~Ruiz de Elvira, B.~Kubis and U.~G.~Mei\ss{}ner,
Phys. Rev. Lett. \textbf{115}, 092301 (2015)
doi:10.1103/PhysRevLett.115.092301
[arXiv:1506.04142 [hep-ph]].

\bibitem{Junnarkar:2013ac}
P.~Junnarkar and A.~Walker-Loud,
Phys. Rev. D \textbf{87}, 114510 (2013)
doi:10.1103/PhysRevD.87.114510
[arXiv:1301.1114 [hep-lat]].

\bibitem{Shifman:1978zn}
M.~A.~Shifman, A.~I.~Vainshtein and V.~I.~Zakharov,
Phys. Lett. B \textbf{78}, 443-446 (1978)
doi:10.1016/0370-2693(78)90481-1

\bibitem{LZ:2022lsv}
J.~Aalbers \textit{et al.} [LZ],
Phys. Rev. Lett. \textbf{131}, no.4, 041002 (2023)
doi:10.1103/PhysRevLett.131.041002
[arXiv:2207.03764 [hep-ex]].

\bibitem{Boughezal:2022nof}
R.~Boughezal, Y.~Huang and F.~Petriello,
Phys. Rev. D \textbf{106}, no.3, 036020 (2022)
doi:10.1103/PhysRevD.106.036020
[arXiv:2207.01703 [hep-ph]].

\bibitem{Ellis:2016jkw}
J.~Ellis,
Comput. Phys. Commun. \textbf{210}, 103-123 (2017)
doi:10.1016/j.cpc.2016.08.019
[arXiv:1601.05437 [hep-ph]].
\end{thebibliography}
\end{document}